\documentstyle[preprint,eqsecnum,aps]{revtex}

\begin{document}

\tightenlines
\draft

%\preprint{SNUTP 98-xxx}

\title{Topological black holes in the dimensionally continued gravity}

\author{Rong-Gen Cai}
\address{Center for Theoretical Physics, Seoul National University,
        Seoul 151-742, Korea}
\author{Kwang-Sup Soh}
\address{Department of Physics Education, Seoul National University,
Seoul 151-742, Korea}

\maketitle

\begin{abstract}
We investigate the topological black holes in a special class of Lovelock 
gravity. In the odd dimensions, the action is the Chern-Simons form for the 
anti-de Sitter group. In the even dimensions, it is the Euler density 
constructed with the Lorentz part of the anti-de Sitter curvature tensor.
The Lovelock coefficients are reduced to two independent parameters: 
cosmological constant and gravitational constant. The event horizons of these
topological black holes may have constant positive, zero or negative  
curvature. Their thermodynamics is analyzed and electrically charged 
topological black holes  are also considered. We emphasize the differences 
due to  the different curvatures of event horizons. As a comparison,
we also discuss the topological black holes in the higher dimensional 
Einstein-Maxwell theory with a negative cosmological constant.

\end{abstract}
\pacs{PACS numbers: 04.20.Jb, 04.20.Gz, 97.60.Lf}

\section{Instruction}
Over the past few years there has been a lot of interest in black holes in
the anti-de Sitter spacetimes. This study is motivated by  the discovery
of Ba\~nados-Teitelboim-Zanelli (BTZ) black holes \cite{BTZ}, which are exact
 solutions in the three-dimensional Einstein gravity with a negative 
cosmological constant, and are locally equivalent to a three-dimensional 
anti-de Sitter space. That is,
the BTZ black holes can be constructed by identifying some discrete points 
along a boost Killing vector in the three-dimensional anti-de Sitter 
space \cite{BTZ1}.
Using such kind of identifications, the so-called constant curvature black 
holes can also be constructed in the four-dimensional \cite{Amin} 
 as well as higher dimensional \cite{Ban} anti-de Sitter spacetimes. 
The Euclidean manifold topologies of these black holes are 
$ R^{D-1}\times S^1$, where $S^1$ is
the topology of event horizons, in contrast to the usual 
topology of black holes $R^2\times S^{D-2}$. Because of the unusual
asymptotic behavior of these constant curvature black holes, however,
 identifying the globally conserved quantities seems difficult 
(For a quasilocal formulation see \cite{Crei}).

On the other hand, except for the Kerr-Newmann-anti-de Sitter black hole,
whose event horizon has the topology $S^2$, in the four-dimensional 
Einstein-Maxwell theory with a negative cosmological constant, it has been  
found recently that there exist black hole solutions whose  event
 horizons may have zero or negative constant curvature and their topologies 
are no longer the two-sphere $S^2$.
Because of the different topological structures of even horizons,
 properties of these black holes are quite different from those of 
black holes with usual spherical topology horizon.
These black holes have been studied  extensively in many aspects such as exact
solutions \cite{Lemos,Huang,Cai1,Klemm1,Klemm}, 
thermodynamics \cite{Brill,Vanzo},
pair production \cite{Mann1}, gravitational collapse \cite{Mann2,Lemos2}, and
others \cite{Cai2,Dann,Cal,Klemm2,And,Die}.

So far, most of the works have been limited in the Einstein gravitational 
theory.  Quite recently, Klemm \cite{Klemm} has found topological black hole 
solutions in the Weyl  conformal gravity.  In a previous paper, we have 
investigated the topological black holes \cite{Cai3} in
a class of dilaton gravity with a Liouville-type dilaton potential. Differing
from the topological black holes in the Einstein-Maxwell theory, which approach 
asymptotically the anti-de Sitter spaces, the topological dilaton black holes
are asymptotically neither the anti-de Sitter spaces nor  de Sitter spaces 
or Minkowski spacetimes. But the negative effective cosmological constant
plays a crucial role in the existence of these black hole solutions, as the
negative cosmological constant does in the Einstein-Maxwell theory.

In the present paper, we would like to investigate the topological 
black holes in the higher dimensional spacetimes. In the Einstein-Maxwell 
theory, the higher dimensional, spherically symmetric black holes have been
 studied by Myers and Perry \cite{Myers}. And their analogues in the 
Brans-Dicke theory have been investigated recently in \cite{Cai4}. Therefore,
 four-dimensional topological black holes  have their natural generalization 
in the higher dimensional Einstein-Maxwell theory  with a negative cosmological
 constant. For example, there are the static topological black holes in 
four-dimensional spacetimes \begin{equation}
\label{In1}
ds^2=- \left( k-\frac{8\pi M}{\omega _2\ r}+\frac{16\pi ^2 Q^2}
      {\omega _2^2\ r^2}+ \frac{r^2}{l^2}\right) dt^2
     + \left( k-\frac{8\pi M}{\omega _2\ r}+\frac{16\pi ^2Q^2}
     {\omega_2^2\  r^2}+ \frac{r^2}{l^2}\right)^{-1}dr^2
      +r^2d\Sigma^2_2,
\end{equation}
where $d\Sigma^2_2$ is the line element of a two-dimensional
hypersurface $\Sigma_2$ with constant curvature $2k$,
\begin{equation}
\label{In2}
d\Sigma_2^2 =\left \{
\begin{array}{ll}
         d\theta^2 +\sin^2\theta d\phi^2 &\ \ {\rm for}\ \ \  k=1, \\
         d\theta^2 +\theta^2d\phi^2      &\ \ {\rm for} \ \ \ k=0, \\
         d\theta^2 +\sinh^2\theta d\phi^2 &\ \ {\rm for}\ \  \ k=-1 .
\end{array} \right.
\end{equation}
Here $M$ and $Q$ are the  mass and charge of the black holes, $-3l^{-2}$ is the
negative cosmological constant, and $\omega _2$ is the area of the horizon 
hypersurface $\Sigma_2$.  In (\ref{In2}), without loss of the generality,
we have  used the coordinates in which the constant curvature of the 
two-dimensional
hypersurface of the event horizon  is $1$, $0$, and $-1$, respectively. 
 When $k=1$, the solution (\ref{In1}) is
just the Reissner-Nordstr\"om-anti-de Sitter black hole spacetime and the event
horizon has the topology $S^2$. When $k=0$, if one identities the coordinates
$\theta$ and $\phi$  with certain periods, the resulting topology of event 
horizon is a torus $T^2$. The event horizon is a hyperbolic surface as $k=-1$. 
Of interest is to note that the event horizon still appears even if the
 mass $M$ is 
negative in that case, and such kind of negative mass black holes might be
formed by the regular gravitational collapse \cite{Mann2}.  
 In addition, because of the different topological structures of event horizons,
their thermodynamic behaviors are  quite different \cite{Brill,Vanzo}.
As a natural extension, we have  topological black hole
solutions in the higher dimensional Einstein-Maxwell theory with a negative
cosmological constant $\Lambda =-(D-1)(D-2)/2l^2$:
\begin{eqnarray}
\label{In3}
ds^2 &=& -\left(k-\frac{16\pi M}{(D-2)\omega _{D-2}\ r^{D-3}}+
     \frac{16\pi ^2 Q^2}{\omega ^2_{D-2}\ r^{2(D-3)}}
     +\frac{r^2}{l^2}\right)dt^2       \nonumber \\      
  &+& \left(k-\frac{16\pi M }{(D-2)\omega _{D-2}\ r^{D-3}}+
      \frac{16\pi ^2 Q^2}{\omega ^2_{D-2}\ r^{2(D-3)}}
       +\frac{r^2}{l^2}\right)^{-1}dr^2 +r^2d\Sigma^2_{D-2},
\end{eqnarray}
where  $d\Sigma^2_{D-2}=\gamma_{mn}dx^mdx^n$ is a (D-2)-dimensional
 hypersurface $\Sigma_{D-2}$  with constant curvature $(D-2)k$, and
 $\omega_{D-2}$ is 
its area. Without loss of generality,  one may normalize 
the constant curvature to $k=1$, $0$, and $-1$, respectively. 
These black holes have
 similar properties as those in the four-dimensional spacetime.   
 For the discussion about the higher dimensional topological
uncharged black holes see \cite{Dann2}.

Instead of the pure Einstein gravity, in this paper, 
 we consider the topological 
black holes in the  so-called dimensionally continued gravity \cite{BTZ2}.
This theory will be reviewed briefly in the next section.  The topological
black hole solutions will be presented and discussed in Sec. III. 
The section IV is devoted to the case including the Maxwell field. 
We summarize our results in Sec.V. In the Appendix we will discuss the 
thermodynamics of the topological black holes (\ref{In3}) in the 
Einstein-Maxwell theory.

\section{Dimensionally continued gravity}

 The dimensionally continued gravity is a special class of the Lovelock 
gravity \cite{Love},  which   may be regarded as the most general 
generalization to higher dimensions of the 
 Einstein gravity. The Lovelock  action is a sum of the dimensionally 
continued Euler characteristics of all dimensions below the spacetime 
dimension $D (\ge 3)$ under consideration. It can be written down as \cite{BTZ2}
\begin{equation}
\label{Dim1}
I=\kappa \sum ^{n}_{p=0}\alpha _pI_p,
\end{equation}
where
\begin{equation}
\label{Dim2}
I_p=\int \epsilon_{a_1\cdots a_D}R^{a_1a_2}\wedge \cdots \wedge 
      R^{a_{2p-1}a_{2p}}\wedge e^{a_{2p+1}}\wedge \cdots \wedge e^{a_D}.
\end{equation}
Here $e^a$ is the local frame one-form, $R^a_{\ b}$ is the curvature two-form
defined as $R^a_{\ b}=dw^a_{\ b} +w^a_{\ c}\wedge w^c_{\ b}$, and
 $w^a_{\ b}$ is the spin connection, $a_i=\{0,1,\cdots,D-1\}$. The coefficients 
$\alpha_p$ are arbitrary
constants with dimensions [length]$^{-(D-2p)}$ and $\kappa$ has units of 
action.

The Lovelock action (\ref{Dim1}) has the advantage which keeps the field 
equations of motion for the metric of second order, as the pure 
Einstein-Hilbert action. But it includes $[D/2]$ arbitrary constants $\alpha_p$,
which makes difficult to extract physical information from the solutions
of the equations of motions. In \cite{BTZ2} a proposal has been suggested 
to reduce
these arbitrary constants to two: a cosmological constant and a 
gravitational constant.  This proposal was made by embedding the Lorentz group
$SO(D-1,1)$ into a larger group, anti-de Sitter group $SO(D-1,2)$. In this way 
the Lovelock theory is divided into two different branches according to the 
spacetime  dimensions: odd dimensions and even dimensions. The coefficients
$\alpha_p$ are  given by
\begin{equation}
\label{alpha}
\alpha_p= \left \{
\begin{array}{ll}
\frac{1}{D-2p}\left(
\begin{array}{c}
n-1    \\
p
\end{array} \right) 
l^{-D+2p}  &   \ \  {\rm for}\ \ D=2n-1 \\
\left( \begin{array}{c}
n\\
p
\end{array} \right) l^{-D+2p}  & \ \ {\rm for}\ \ D=2n,
\end{array} \right.
\end{equation}
where $l$ is a length.

In the odd dimensions, the Lagrangian ${\cal L}_{2n-1}$ is
\begin{equation}
\label{oddlag}
{\cal L}_{2n-1}=\kappa \sum^{n-1}_{0} \alpha_p   \epsilon _{a_1\cdots a_D}
          R^{a_1a_2}\wedge \cdots \wedge R^{a_{2p-1}a_{2p}}
      \wedge e^{a_{2p+1}}\wedge \cdots\wedge e^{a_D}.
\end{equation}
For later convenience, the units are chosen so that
\begin{equation}
\label{kappa1}
\kappa =\frac{l}{(D-2)!\omega_{D-2}} \ \ \ \ {\rm for }\ \  D=2n-1,
\end{equation}
where $\omega_{D-2}$ is the area of a $(D-2)$-dimensional hypersurface
$\Sigma_{D-2}$  which will be defined later.
In the even dimensions, the Lagrangian ${\cal L}_{2n}$ is given by
\begin{equation}
\label{evenlag}
{\cal L}_{2n}= \kappa (R^{a_1a_2} +l^{-2}e^{a_1}\wedge e^{a_2})\wedge \cdots 
             \wedge (R^{a_{D-1}a_{D}}+l^{-2}e^{a_{D-1}}\wedge e^{a_D})
             \epsilon _{a_1a_2\cdots a_D},
\end{equation}
where we choose the units so that
\begin{equation}
\label{kappa2}
\kappa=\frac{l^2}{2D(D-2)!\omega_{D-2}}\ \ \ \ {\rm for}\ \  D=2n.
\end{equation}
In the cases $D=4$ and $D=3$, the two Lagrangians reduce to that of 
the Einstein gravity with a negative cosmological constant. For the details
of the construction of the two Lagrangians see \cite{BTZ2}.

Correspondingly, the equations of motion from (\ref{oddlag}) 
and (\ref{evenlag}) can be derived as 
\begin{equation}
\label{oddeq}
   (R^{a_1a_2} + l^{-2}e^{a_1}\wedge e^{a_2})\wedge \cdots 
      \wedge (R^{a_{2n-3}a_{2n-2}}+l^{-2}e^{2n-3}\wedge e^{2n-2})
          \epsilon _{a_1a_2\cdots a_{2n-1}}=0
\end{equation}
in the odd dimensions ($D=2n-1$), and
\begin{equation}
\label{eveneq}
  ( R^{a_1a_2} + l^{-2}e^{a_1}\wedge e^{a_2})\wedge \cdots 
    \wedge (R^{a_{2n-3}a_{2n-2}}+l^{-2}e^{2n-3}\wedge e^{2n-2})
          \wedge e^{2n-1}\epsilon _{a_1a_2\cdots a_{2n}}=0
\end{equation}
in the even dimensions ($D=2n$), from which it is easy to see that 
the anti-de Sitter space is a special solution to these equations 
of motion.

In \cite{BTZ2}, the static, spherically symmetric black hole solutions
 are obtained.  
The metric of the black hole solutions is 
 \begin{equation}
\label{k=1}
ds^2=-g^2(r) dt^2 +g^{-2}(r)dr^2 +r^2d\Omega^2,
\end{equation}
where
\begin{equation}
\label{k12}
g^2=\left \{
\begin{array}{ll}
     1-(2M/r)^{\frac{1}{n-1}}+(r/l)^2  &\ \ \ {\rm for}\ \ D=2n \\
     1-(M+1)^{\frac{1}{n-1}}+(r/l)^2   &\ \ \ {\rm for}\ \ D=2n-1,
\end{array} \right.
\end{equation}
$M$ is the mass of the hole and $d\Omega ^2$ is the metric on the
 unit ($D-2$)-sphere. Although this black hole solution (\ref{k=1}) has
different quantum properties from the higher dimensional
 Schwarzschild-anti-de Sitter black hole, its  Euclidean topology is still
$R^2\times S^{D-2}$, where $S^{D-2}$ is the topology of its event horizon.
That is, its event horizon is a ($D-2$)-dimensional sphere.
 In this  work,  we pay attention to the black holes
whose event horizons are $(D-2)$-dimensional hypersurfaces with
constant curvature which may  be positive, zero or  negative, and hence 
the topology of event horizon is no longer the $(D-2)$-dimensional sphere 
$S^{D-2}$.    Here we should mention that, in this dimensionally
continued  gravity considered above, the Oppenheimer-Snyder gravitational 
collapse in the case of even dimensions has been studied recently by 
Ilha and Lemos \cite{Ilha}, it has been found that the even dimensional
black holes (\ref{k=1}) emerge as the final state of regular dust fluid. 
The wormhole solutions have been found in \cite{Li}.

\section{Topological black holes and Thermodynamics }

In this section we discuss topological black hole solutions to the equations 
of motion (\ref{oddeq}) and (\ref{eveneq}). The event horizon of 
these topological black holes is a $(D-2)$-dimensional hypersurface with
constant curvature. The topology of horizon may be sphere, torus or other
higher genus Riemann surfaces.

\subsection{Static solutions and general consideration of thermodynamics}

In order to obtain simplified equations of motion, it turns out that it 
is more convenient to work in the Hamiltonian form \cite{BTZ2}. The
 Hamiltonian formulation
of the Lovelock action (\ref{Dim1}) has been provided in \cite{TZ}. 
 The Hamiltonian constraint is
\begin{equation}
\label{ham1}
{\cal H}=-\sqrt{\det (h_{ij})}\sum^{n-1}_{p=0}\frac{D-2p}{2^p}\alpha_p
        \delta ^{[i_1\cdots i_{2p}]}_{[j_1\cdots j_{2p}]}
       \tilde{R}^{j_1j_2}_{i_1i_2}\tilde{R}^{j_3j_4}_{i_3i_4}\cdots 
         \tilde{R}^{j_{2p-1}j_{2p}}_{i_{2p-1}i_{2p}},
\end{equation}
where $\tilde{R}^{ij}_{kl}$ are the spatial components of the Riemann 
tensor. They depend on the velocities through the Gauss-Codazzi equations 
\begin{equation}
\tilde{R}_{ijkl}=R_{ijkl}+K_{ik}K_{jl}-K_{il}K_{jk},
\end{equation}
where $R_{ijkl}$ are the components of the intrinsic curvature tensor of
 the spatial sections and $K_{ij}$ is the second fundamental form defined 
as $K_{ij}= \frac{1}{2N^{\bot}}(-\dot{h}_{ij}+N_{i;j}+N_{j;i})$, where $h_{ij}$,
$ N^{\bot}$, and $N^i$ are the reduced metric, lapse function and shift vectors 
in the standard Arnowitt-Deser-Misner (ADM) decomposition of spacetime.

     We are looking for the static topological black hole solutions. 
So the metric are assumed as
\begin{equation}
\label{metric}
ds^2 =-N^2(r)g^2(r) dt^2 +g^{-2}(r)dr^2 +r^2 d\Sigma^2_{D-2},
\end{equation}
where $d\Sigma_{D-2}^2=\gamma_{mn}(x)dx^mdx^n$ is the metric of 
$(D-2)$-dimensional hypersurface $\Sigma_{D-2}$ with constant curvature 
 $(D-2)k$ and its area is denoted by 
$\omega_{D-2}$, which is just the one in (\ref{kappa1}) and (\ref{kappa2}).
 The functions $N^2(r)$ and $g^2(r)$ need to be determined.
 In the metric (\ref{metric}) the nonvanishing spatial components of 
 curvature tensors are
\begin{eqnarray}
\label{tensor}
&&¤R ^{m_1m_2}_{n_1n_2}=\frac{f(r)}{r^2}\delta ^{[m_1m_2]}_{[n_1n_2]}, 
            \nonumber   \\
&& R^{rm}_{rn}=\frac{f'(r)}{2r} \delta ^m_n,
\end{eqnarray}
where  a prime denotes the derivative with respect to $r$ and $f(r)=k-g^2(r)$.
 Substituting (\ref{tensor}) into the Hamiltonian 
constraint (\ref{ham1}) yields 
\begin{equation}
\label{ham2}
{\cal H}=-(D-2)!\sqrt{\gamma}g^{-1}\left [r^{D-1}\sum^{n-1}_{p=0}
       \alpha_p (D-2p)\left(\frac{k-g^2}{r^2}\right)^p\right]'.
\end{equation}
Using the coefficients (\ref{alpha}) and units (\ref{kappa1}) and
 (\ref{kappa2}), one has the action
\begin{equation}
\label{action2}
I=(t_2-t_1)\int dr \ NF'(r) +B,
\end{equation}
where $B$ stands for a surface term and the function $F$ is given by
\begin{equation}
\label{Ffunction}
       F[r,g(r)]=\left \{
\begin{array}{ll}
      \frac{1}{2}r[ k+(r/l)^2-g^2(r)]^{n-1}  & \ \ {\rm for} \ \ D=2n, \\
          {}    [k+(r/l)^2-g^2(r)]^{n-1}  & \ \ {\rm for} \ \ D=2n-1.
\end{array} \right.
\end{equation}
Varying the action (\ref{action2}) with respect to $N$ and $F$, one has the 
equations of motion
\begin{equation}
\label{eqmo}
\frac{dF}{dr}=0, \ \ \ \frac{dN}{dr}=0,
\end{equation}
which have solutions
\begin{eqnarray}
&& F[r,g(r)]=\tilde {C}\equiv {\rm Const}, \nonumber\\
&& N(r)=N_{\infty} \equiv {\rm Const}.
\end{eqnarray}
Here the integration constant $N_{\infty}$ can be taken to be one by rescaling
the time coordinate $t$. And the constant $\tilde{C}$ is related to, up to an 
additive constant, the mass $M$  of black holes,
\begin{equation}
\label{mass}
\tilde{C}=M +\tilde{C}_0.
\end{equation} 
This additive constant $\tilde{C}_0$ is achieved by choosing an appropriate 
reference background.
Thus we obtain the metric function $g^2(r)$: 
\begin{equation}
\label{g^2}
g^2(r)=\left \{
\begin{array}{ll}
k-\left(\frac{2M+2\tilde{C}_0}{r}\right)^{\frac{1}{n-1}}
        +\left(\frac{r}{l}\right)^2 
        &\  \ {\rm for}\ \ D=2n, \\
k-(M+\tilde{C}_0)^{\frac{1}{n-1}}+\left(\frac{r}{l}\right)^2
         & \ \ {\rm for}\ \ D=2n-1.
\end{array} \right.
\end{equation}
The mass for the even dimensional solutions has the dimensions of length,
and is dimensionless for the odd dimensional solutions. This is because
we have chosen the different units (\ref{kappa1}) and (\ref{kappa2}).
 To analyze the singularities of the solutions (\ref{g^2}), let us write down 
some curvature invariants
\begin{equation}
\label{curvature}
R=\left \{
\begin{array}{ll}
-\frac{D(D-1)}{l^2}+\left[\frac{2D}{(D-2)^2}+D^2 -5D +2\right]
  \frac{1}{r^2}\left(\frac{2M+2\tilde{C}_0}{r}\right)^{\frac{1}{n-1}}
     & \ \ {\rm for}\ \ D=2n, \\
-\frac{D(D-1)}{l^2} +\frac{(D-2)(D-3)}{l^2}(M+\tilde{C}_0)^{\frac{1}{n-1}}
      & \ \ {\rm for} \ \ D=2n-1.
\end{array} \right.
\end{equation}
\begin{eqnarray}
R_{\alpha\beta}R^{\alpha\beta} &=&
   \frac{1}{2}[(g^2)'']^2 +\frac{(D-2)}{r}(g^2)''(g^2)' 
  +\frac{D(D-2)}{2r^2}[(g^2)']^2 
     \nonumber \\
  &-& \frac{2(D-2)(D-3)}{r^3}(k-g^2)(g^2)'
     +\frac{(D-2)(D-3)}{r^4}(k-g^2)^2.
\end{eqnarray}
\begin{equation}
R_{\alpha\beta\gamma\delta}R^{\alpha\beta\gamma\delta}
   =[(g^2)'']^2 +\frac{2(D-2)}{r^2}[(g^2)']^2 +\frac{2(D-2)(D-3)}{r^4}
      (k-g^2)^2.
\end{equation}

Before proceeding to discuss the solution (\ref{g^2}) for different 
curvature $k$, let us consider generally 
 the thermodynamics of black hole solutions (\ref{metric}). Assuming the metric
(\ref{metric}) describes a black hole with  event horizon at $r_+$, one has a 
black hole with event horizon being a $(D-2)$-dimensional
 hypersurface $\Sigma_{D-2}$
whose curvature  may be positive, zero or negative. Apart from the topology 
structure of horizons, to determine the Hawking temperature of the hole, 
it turns
out that it is convenient to continue the metric (\ref{metric}) to its Euclidean
manifold with the Euclidean time $\tau=it$,
\begin{equation}
\label{emetric}
ds^2=N^2(r)g^2(r)d\tau^2 +g^{-2}(r)dr^2 +r^2d\Sigma_{D-2} ^2.
\end{equation}
For an arbitrary period of the Euclidean time $\tau$, there is a conical 
singularity
at the black hole horizon. To remove this singularity in the Euclidean manifold 
(\ref{emetric}), one has to take a special  period whose inverse just gives 
the Hawking temperature of black holes
\begin{equation}
\label{tem}
T=\left. \frac{(N^2g^2)'}{4\pi N}\right |_{r=r_+}.
\end{equation}
For the solutions (\ref{g^2}), Using (\ref{tem}) yields the Hawking temperature
of the holes 
\begin{equation}
\label{temtot}
T=\left \{
\begin{array}{ll}
\frac{k+(2n-1)(r_+/l)^2}{4\pi (n-1)r_+} &\ \ {\rm for}\ \ D=2n, \\
\frac{r_+}{2\pi l^2} &\ \ {\rm for}\ \ D=2n-1.
\end{array} \right.
\end{equation}

Usually the black hole entropy satisfies the so-called area formula.
 That is, the black hole entropy equals to one quarter of event 
horizon area. But this formula does not always hold. It has been proved 
that this formula holds only for the Einstein 
gravity and in fact, black hole entropy comes from a surface term of 
gravitational action at the horizon 
\cite{Cai4,Jaco,BTZ3}. That is,  the black hole entropy is related to 
the gravitational theory under consideration. To get the black hole entropy, 
there are several methods 
available now.  Here we adopt a simpler method for this goal.
This method is based on  the fact that, as  thermodynamic systems, 
black holes  must obey the first law of thermodynamics
\begin{equation}
\label{first}
dM=TdS+ \sum_{i=1} \mu _i Q_i,
\end{equation}
where $M$  is the mass of black  holes,  $T$ and $S$ are the Hawking 
temperature and Hawking-Bekenstein entropy of the black holes,
 respectively. $\mu_i$ are
the chemical potentials corresponding to the conserved charges $Q_i$.
 Using (\ref{first}) one has
\begin{eqnarray}
\label{entropy1}
S &=&\int T^{-1}dM +S_0 \nonumber \\
  &=& \int T^{-1}\left(\frac{\partial M}{\partial r_+}\right)_{Q_i}dr_+ +S_0.
\end{eqnarray}
 Here it should be reminded that, in the 
integration (\ref{entropy1}), the charges $Q_i$ should be taken as constants. 
$S_0$ is an integration constant, which  can 
be fixed by using the argument that the black hole entropy 
should vanish as the event
 horizon of black holes disappears. Therefore, the expression (\ref{entropy1}) can 
be rewritten as  
\begin{equation}
\label{entropy}
S=\int ^{r_+}_{0}T^{-1}\left( \frac{\partial M}{\partial r_+}\right)_{Q_i} dr_+.
\end{equation}
Thus, once given the Hawking temperature and the mass expressed by the horizon radius
$r_+$ and charges $Q_i$,
one can obtain the entropy of the black hole and needs not to know in which 
gravitational theory  the black hole solutions are. 

 For the solutions (\ref{g^2}), the charges $Q_i$ are absent. Using the Hawking
temperature (\ref{temtot}), we obtain the black hole entropy of solutions (\ref{g^2})
\begin{eqnarray}
\label{entropytot1}
S &=&2\pi (n-1)\int^{r_+}_0 r_+\left [k+\left(\frac{r_+}{l}\right )^2
           \right ]^{n-2}dr_+ \nonumber\\
  &=& \pi l^2\left \{\left [k+\left (\frac{r_+}{l}\right )^2\right ]^{n-1}-k
             \right\},
\end{eqnarray}
in the even dimensions, and
\begin{eqnarray}
\label{entropytot2}
S &=& 4\pi (n-1) \int^{r_+}_0 \left [k+\left(\frac{r_+}{l}\right )^2
      \right ]^{n-2}dr_+ \nonumber \\
  &=&4\pi (n-1) \int ^{r_+}_0 \sum^{n-2}_{m=0}\left ( 
\begin{array}{c}
n-2  \\ m
\end{array} \right )
\left(\frac{r_+}{l}\right)^{2m}k^{n-2-m} dr_+ \nonumber\\
&=& 4\pi (n-1) l \sum^{n-2}_{m=0}\left(
\begin{array}{c}
 n-2 \\ m 
\end{array} \right) \frac{1}{2m+1}\left(\frac{r_+}{l}\right)^{2m+1}k^{n-2-m}.
\end{eqnarray}
in the odd dimensions. Obviously, they do not obey the usual area formula.

To see the stability of black holes against the Hawking radiation, it is useful
to compute the heat capacity defined as 
$C_{Q_i}\equiv (\partial M/\partial T)_{Q_i}$. For the solutions (\ref{g^2}),
we obtain
\begin{equation}
\label{heattot}
C=\left \{
\begin{array}{ll}
2\pi (n-1)r_+^2\frac{[k+(r_+/l)^2]^{n-2}[(2n-1)(r_+/l)^2+k]}
    {(2n-1)(r_+/l)^2-k}& \ \ {\rm for}\ \ D=2n, \\
4\pi (n-1)r_+[k+(r_+/l)^2]^{n-2}& \ \ {\rm for}\ \ D=2n-1.
\end{array} \right.
\end{equation}
Here we would like to stress that these physical quantities (\ref{temtot}),
(\ref{entropytot1}), (\ref{entropytot2}) and (\ref{heattot}) are all 
expressed in terms of
the horizon radius $r_+$ and curvature $k$, the addition constant $\tilde{C}_0$ 
does not explicitly enter these quantities.

We now turn to discussing the solutions (\ref{g^2}) for different horizon
curvature $k$. As mentioned in the Introduction, without loss of generality, 
the curvature  can be normalized as $k=1$, $0$ and $-1$, respectively. 
In the case $k=1$ and horizon is spherical topology, this solution 
has  already been analyzed in some detail in \cite{BTZ2}. Just as pointed out by
Birmingham \cite{Dann2}, however, it should be noticed that in the higher 
dimensions, even in the case $k=1$, there  still exist the possibilities of
non-spherical topology for the horizon hypersurface. In addition, to
 compare with
the other two cases, we also summarize and comment the $k=1$ black hole
 solutions below.

\subsection{ $k=1$ solutions} 
 
To analyze the solutions (\ref{g^2}), one has to first fix the additive constant 
$\tilde{C}_0$.  In \cite{BTZ2},  Ba\~nados, Teitelboim, and Zanelli 
 used a criterion to fix the constant $\tilde{C}_0$ in (\ref{mass}) 
that for zero energy the horizon should disappear. 
They fixed the constant $\tilde{C}_0$ as
\begin{equation}
\label{C1}
\tilde{C}_0=\left \{
\begin{array}{ll}
0 & \ \  \ {\rm for }\ \ D=2n, \\
1 & \ \ \  {\rm for }\ \ D=2n-1.
\end{array} \right.
\end{equation}
If use this choice, we have the solutions 
\begin{equation}
\label{solution1}
ds^2=-[1-(2M/r)^{\frac{1}{n-1}}+(r/l)^2]dt^2 +[1-(2M/r)^{\frac{1}{n-1}}
       +(r/l)^2]^{-1}dr^2 +r^2d\Sigma_{D-2}^2,
\end{equation}
in the even dimensions $D=2n$, and 
\begin{equation}
\label{solution2}
ds^2=-[1-(M+1)^{\frac{1}{n-1}}+(r/l)^2]dt^2 +[1-(M+1)^{\frac{1}{n-1}}
        +(r/l)^2]^{-1}dr^2 +r^2d\Sigma_{D-2}^2,
\end{equation}
in the odd dimensions $D=2n-1$. When $D=4$ and $D=3$, the solutions 
(\ref{solution1}) and (\ref{solution2}) reduce, respectively, to the
Schwarzschild-anti-de Sitter solution and BTZ black hole solution \cite{BTZ}.
For the solution (\ref{solution1}), the zero mass reference background is the
$D=2n$ dimensional anti-de Sitter space. For the solution (\ref{solution2}),
the zero mass reference background describes  in fact a zero mass black hole,
the anti-de Sitter space is recovered as $M=-1$, just as happens in the BTZ black
 holes \cite{BTZ}. Here it is worth noting that, in the superstring theory,
zero mass BTZ black hole and $M=-1$ anti-de Sitter space are both ground states,
but in the different sectors. The anti-de Sitter space is the ground state in 
the NS-NS sector, while the zero mass BTZ black hole in the R-R sector 
\cite{Henx,Strom}.

Both of the solutions (\ref{solution1}) and (\ref{solution2}) approach the 
anti-de Sitter space, and a scalar singularity exists at $r=0$ 
(This singularity does not exit in the dimension $D=3$ for arbitrary mass).
This singularity may be covered by an event horizon, which is determined by
the equation $g^2(r)=0$, that is,
\begin{equation}
\left \{
\begin{array}{ll}
1-(2M/r_+)^{\frac{1}{n-1}} +(r_+/l)^2=0 & \ \ {\rm for}\ \ D=2n,\\
1-(M+1)^{\frac{1}{n-1}}+(r_+/l)^2=0 & \ \  {\rm for } \ \ D=2n-1.
\end{array} \right.
\end{equation}
In the even dimensions, one cannot get generally  an explicit 
expression of horizon in
terms of the mass of the hole. However, in the odd dimensions, one has
\begin{equation}
r_+=l\sqrt{(M+1)^{\frac{1}{n-1}}-1}.
\end{equation}
In spite of the dimension, both the solutions (\ref{solution1}) and 
(\ref{solution2}) have only one horizon. In the even dimensions, the causal
structure is similar to that of the Schwarzschild-anti-de Sitter black hole, 
while it is similar to that of the BTZ black hole in the odd dimensions. 
 For both cases, the Penrose
diagrams are exhibited in {\cite{BTZ2}.

As $k=1$, the Hawking temperature (\ref{temtot}) reduces to
\begin{equation}
\label{tem1}
T=\left \{
\begin{array}{ll}
\frac{1+(2n-1)(r_+/l)^2}{4\pi (n-1)r_+}& \ \ {\rm for } \ \ D=2n, \\
\frac{r_+}{2\pi l^2}&\ \ {\rm for}\ \  D=2n-1.
\end{array} \right.
\end{equation}
Obviously, the behavior of the Hawking radiation is quite different because of
the dimension. In the odd dimensions, $T\rightarrow 0$ as $r_+\rightarrow 0$,
while $T\rightarrow \infty$  in the even dimensions.
To see clearly  this behavior and the stability of black holes against the 
Hawking radiation, let us write down the heat capacity.
From (\ref{heattot}), we have
\begin{equation}
\label{heat1}
C=\left \{
\begin{array}{ll}
  2\pi (n-1)r_+^2\left[1+\left(\frac{r_+}{l}\right)^2\right]^{n-2}
 \frac{[(2n-1)(r_+/l)^2+1]}{[(2n-1)(r_+/l)^2-1]} & \ \ \ {\rm for }\ \ D=2n, \\
4\pi (n-1)r_+\left[1+\left(\frac{r_+}{l}\right)^2\right]^{n-2} &\ \ \ {\rm for}
          \ \ D=2n-1.
\end{array} \right.
\end{equation}
From the heat capacity, we see that it is always positive in the 
  odd dimensions.  Therefore,  the odd
dimensional black holes can be in thermal equilibrium with Hawking radiation with
arbitrary volume, as the BTZ black hole.  For the even dimensional black holes,
 however,
the heat capacity is negative as $r_+< l/\sqrt{2n-1}$, positive as $r_+>l/ \sqrt{2n-1}$.
That is, there is a transition point for the even dimensional black holes at
 $r_+=l/\sqrt{2n-1}$, thereby the heat capacity suffers from an infinite jump.

The entropies of the  wo kinds of black holes, from (\ref{entropytot1})
 and (\ref{entropytot2}),  are 
\begin{equation}
S=\left \{
\begin{array}{ll}
\pi l^2\left \{\left[1+\left(\frac{r_+}{l}\right)^2\right]^{n-1}-1\right\} 
  &\ \ {\rm for}\ \ D=2n, \\
4\pi (n-1)l \sum ^{n-2}_{m=0} \left (
\begin{array}{c}
 n-2 \\ m
\end{array}\right)
\frac{1}{2m+1}\left(\frac {r_+}{l}\right)^{2m+1} & \ \ {\rm for}\ \ D=2n-1.
\end{array}
\right.
\end{equation}

As a discussion on the choice of the additive constant $\tilde{C}_0$ 
in (\ref{C1}), we note that the solution (\ref{solution2}) reduces to 
the BTZ black hole  as $D=3$ and $\tilde{C}_0=1$.  However, for 
three-dimensional black holes, the horizon is a circle and its curvature
$k$ must vanish, i.e., $k=0$,  which belongs to the class of solutions discussed
in the next subsection. As a result, we have another choice of $\tilde{C}_0$
for the $(D>3)$  odd dimensional black holes with $k=1$. That is, one may choose
$\tilde{C}_0=0$ as in the even dimensions, and the metric then becomes 
\begin{equation}
ds^2=-[1-M^{\frac{1}{n-1}}+(r/l)^2]dt^2 
      +[1-M^{\frac{1}{n-1}}+(r/l)^2]^{-1}dr^2 +r^2d\Sigma^2_{D-2},
\end{equation}
where  $D>3$. In that case, the vacuum background, as in the even 
dimensions, is still
the anti-de Sitter space. Note that this choice also satisfies the criterion 
that the horizon disappears for zero mass solutions.

\subsection{$k=0$ solutions}

In the case of $k=0$, we fix the constant $\tilde{C}_0$ as
\begin{equation}
\tilde{C}_0=0,  \ \ \ {\rm for\ arbitary\ dimensions}.
\end{equation}
Then we have the solutions  
\begin{equation}
\label{solution3}
ds^2=-[-(2M/r)^{\frac{1}{n-1}} +(r/l)^2]dt^2 
    +[-(2M/r)^{\frac{1}{n-1}}+(r/l)^2]^{-1}dr^2 +r^2d\Sigma^2_{D-2},
\end{equation}
in the even dimensions, and 
\begin{equation}
\label{solution4}
ds^2 =-[-M^{\frac{1}{n-1}}+(r/l)^2]dt^2 
       +[-M^{\frac{1}{n-1}}+(r/l)^2]^{-1}dr^2 +r^2d\Sigma^2_{D-2},
\end{equation}
in the odd dimensions. When $D=3$, the solution (\ref{solution4}) is
the    BTZ black hole solution. 
 We refer to the ground states of these two kinds of 
black hole solutions   as  zero mass black holes, because the 
zero mass solutions in (\ref{solution3}) and (\ref{solution4}) describes 
 a spacetime whose singularity coincides with the event horizon at $r=0$, 
as the zero mass BTZ solution \cite{Henx}.
For the solutions (\ref{solution3}) and (\ref{solution4}), they are both 
asymptotically  locally equivalent to the anti-de Sitter spaces.
The event  horizons are
\begin{equation}
r_+=\left \{
\begin{array}{ll}
l(2M/l)^{\frac{1}{2n-1}} & \ \ {\rm for}\ \ D=2n, \\
lM^{\frac{1}{2n-2}}  &\ \ {\rm for}\ \ D=2n-1.
\end{array}  \right.
\end{equation}
Note from (\ref{curvature}) that the curvature $k$ does not affect the 
singularity of the solution. Therefore their causal structures are similar
 to those of solutions for $k=1$.
According to (\ref{temtot}), the Hawking temperatures are
\begin{equation}
T=\left \{
\begin{array}{ll}
\frac{(2n-1)r_+}{4\pi (n-1)l^2}=\frac{2n-1}{4\pi (n-1) l}(2M/l)^{\frac{1}{2n-1}}
          & \ \ {\rm for}\ \ D=2n, \\
\frac{r_+}{2\pi l^2}=\frac{1}{2\pi l}M^{\frac{1}{2n-2}}  & \ \ {\rm for}\  \ 
          D=2n-1.
\end{array} \right.
\end{equation}
Both of the Hawking temperatures approach zero as the mass goes to zero. Therefore the heat capacity
should be positive. Indeed, from (\ref{heattot}), one has
\begin{equation}
C= \left \{
\begin{array}{ll}
2\pi (n-1) r_+^2 (r_+/l)^{2n-4} & \ \ {\rm for} \ \ D=2n, \\
4\pi (n-1) r_+ (r_+/l)^{2n-4} & \ \ {\rm for} \ \ D=2n-1,
\end{array} \right.
\end{equation}
which are always positive.  For the even dimensional black holes, 
from (\ref{entropytot1}), the entropy is
\begin{equation}
S=\pi l^2 (r_+/l)^{2n-2}.
\end{equation}
When $k=0$, only does   the term $m=n-2$ have 
 the contribution to the entropy in (\ref{entropytot2}).
The entropy of the odd dimensional black hole therefore is
\begin{equation}
S=\frac{4\pi(n-1) l}{2n-3}\left(\frac{r_+}{l}\right)^{2n-3}.
\end{equation}

\subsection{$k=-1$ solutions}

In the case of $k=-1$, we also fix the constant $\tilde{C}_0=0$  
for both solutions.
 The black hole solutions then are 
\begin{equation}
\label{solution5}
ds^2=-[-1-(2M/r)^{\frac{1}{n-1}}+(r/l)^2]dt^2 
     +[-1-(2M/r)^{\frac{1}{n-1}}+(r/l)^2]^{-1}dr^2 +r^2 d\Sigma^2_{D-2},
\end{equation}
in the even dimensions, and 
\begin{equation}
\label{solution6}
ds^2=-[-1-M^{\frac{1}{n-1}}+(r/l)^2]dt^2 
      +[-1-M^{\frac{1}{n-1}}+(r/l)^2]^{-1} dr^2 +r^2d\Sigma^2_{D-2},
\end{equation}
in the odd dimensions ($D>3$). Both of the two solutions asymptotically
approach the anti-de Sitter spaces.  For both cases, the zero mass solution 
is
\begin{equation}
\label{zero}
ds^2=-[-1+(r/l)^2]dt^2 +[-1+(r/l)^2]^{-1}dr^2 +r^2 d\Sigma^2_{D-2},
\end{equation}
from which one can see clearly that the zero mass solution is a 
black hole solution with horizon at $r_+=l$. Using (\ref{tem}),
 the Hawking temperature of the solution is found to be
\begin{equation}
T=\frac{1}{2\pi l}.
\end{equation}
With the help of (\ref{entropytot1}) and (\ref{entropytot2}), the entropy of 
the zero mass black holes is
\begin{equation}
\label{zeroentropy}
S=\left \{
\begin{array}{ll}
\pi l^2 & \ \  {\rm for} \ \ D=2n, \\
4\pi (n-1)l\sum ^{n-2}_{m=0} \left (
\begin{array}{c}
 n-2 \\ m 
\end{array}\right )
\frac{(-1)^{n-2-m}}{2m+1} & \ \ {\rm for}\ \ D=2n-1.
\end{array} \right.
\end{equation}
From the above, one can find that, for the same class of solutions (\ref{zero}),
 when embedded in different gravities, 
the  same black hole solution has the same Hawking temperature, but different
 entropy formula. but, this is not  surprising. This is because black hole 
entropies are
related to the gravitational theories under consideration and come from
a surface term of gravitational action. So they are different for 
different gravitational theories.
An explicit example is that the entropy of the BTZ black hole is 
proportional to the length of outer horizon $r_+$ in
the Einstein theory, but to the length of inner horizon in the
 topological gravity \cite{RB}.
 Note that the zero mass solution (\ref{zero}) is also a special solution
in the Einstein-Maxwell theory (\ref{In3}). We will show in the Appendix that 
the entropy of black holes (\ref{In3}) obey the area formula.
 That is, the entropy of zero mass black hole in (\ref{In3}) is 
also different from the one  in (\ref{zeroentropy}).

In the general case, i.e., $M\neq 0$, we cannot get an explicit expression of 
the horizon in terms of the mass for the even dimensional black holes, but
\begin{equation}
r_+=l\sqrt{1+M^{\frac{1}{n-1}}},
\end{equation}
for the odd dimensional black holes. According to  (\ref{temtot}),
 the Hawking temperatures are
\begin{equation}
T=\left \{
\begin{array}{ll}
\frac{(2n-1)(r_+/l)^2-1}{4\pi (n-1)l^2} & \ \ {\rm for}\ \ D=2n, \\
\frac{r_+}{2\pi l^2}=\frac{1}{2\pi l}\sqrt{1+M^{\frac{1}{n-1}}} & \ \ 
{\rm for} \ \ D=2n-1.
\end{array}\right.
\end{equation}
From (\ref{heattot}), one has the heat capacity
\begin{equation}
\label{ceven}
C=\left \{
\begin{array}{ll}
2\pi (n-1)r_+^2 \left[\left(\frac{r_+}{l}\right )^2-1\right ]^{n-2}
     \frac{(2n-1)(r_+/l)^2-1}{(2n-1)(r_+/l)^2+1}
      & \ \ {\rm for}\ \ D=2n, \\
4\pi (n-1) r_+ [(r_+/l)^2-1]^{n-2} & \ \ {\rm for} \ \ D=2n-1.
\end{array}\right.
\end{equation}
Note from (\ref{solution5}) that $r_+\ge l$, the heat capacity (\ref{ceven}) is 
therefore always positive.  The entropies of the black holes  are
\begin{equation}
S=\left \{
\begin{array}{ll}
\pi l^2 \left [\left(\frac{r_+}{l}\right)^2 -1\right]^{n-1}+\pi l^2
         & \ \ {\rm for}\ \ D=2n, \\
4\pi (n-1)l \sum^{n-2}_{m=0} \left (
\begin{array}{c}
 n-2 \\ m 
\end{array} \right) 
\frac{(-1)^{n-2-m}}{2m+1}\left(\frac{r_+}{l}\right)^{2m+1} & 
         \ \ {\rm for}\ \ D=2n-1.
\end{array} \right.
\end{equation}

For the $k=-1$ solutions, there also exist the so-called negative 
mass black holes when $n=2\tilde{k} +2$ $(\tilde{k}\in {\cal Z})$. But, there
is a critical value, beyond which the singularity at $r=0$ will be naked. 
The critical mass is
\begin{equation}
M_{\rm c}=-\frac{l}{2\sqrt{2\tilde{k}+2}}\left[
     \frac{2\tilde{k}+1}{2\tilde{k}+2}\right]^{2\tilde{k}+1},
\end{equation}
for the solution (\ref{solution5}), and
\begin{equation}
M_{\rm c}=-1,
\end{equation}
for the solution ({\ref{solution6}).  Inspecting (\ref{ceven}),
it is easy to see that the heat capacity is still positive for these
negative mass black holes.

\section{ Charged topological black holes}

Similar to the static, spherically symmetric black holes \cite{BTZ2}, 
the electric charge can also be incorporated to the topological black
 holes discussed in the previous section. The Hamiltonian action of the 
Maxwell field in a curved spacetime is 
\begin{equation}
\label{em}
I_{\rm em}=\int dt\int d^{D-1}x \left[p^i\dot{A}_i-\frac{1}{2}N^{\bot}\left( 
     \beta h^{-1/2}p^ip_i +\frac{h^{1/2}}{2\beta}F^{ij}F_{ij}\right)
     +\varphi p^i_{,i}\right] +B_{\rm em},
\end{equation}
where $N^{\bot}$ is the lapse function, $h$ is the determinant of 
the induced metric of the ADM decomposition of spacetime. $p^i$ is the 
momentum conjugate to the spatial components of the gauge field $A_i$,
 $\varphi=A_0$, and $B_{\rm em}$ is a surface term depending on the boundary 
condition. The constant $\beta$ related 
to chosen units may be taken conveniently 
to be the area of the hypersurface $\Sigma_{D-2}$. For static, electrically charged
black holes in the metric (\ref{metric}), the action (\ref{em})
can be reduced to \cite{BTZ2} 
\begin{equation}
\label{em2}
I_{\rm em}=(t_2-t_1)\int dr \left[-\frac{1}{2}Nr^{D-2}p^2 
      +\varphi (r^{D-2}p)'\right] +B_{\rm em},
\end{equation} 
where 
\begin{equation}
p=\frac{\beta \ p^r}{\sqrt{\gamma}\ r^{D-2}}.
\end{equation}
Combining (\ref{em2}) and (\ref{action2}), one has the reduced action of the
system 
\begin{equation}
\label{actiontot}
I=(t_2-t_1)\int dr [N(F'-\frac{1}{2}r^{D-2}p^2)+\varphi (r^{D-2}p)'] +\tilde{B},
\end{equation}
where $F$ is still given by (\ref{Ffunction}) and $\tilde{B}$ denotes another 
surface term. Varying the action (\ref{actiontot}) with respect to $N$, $g$, $p$,
and $\varphi$, respectively, one has the equations of motion
\begin{eqnarray}
 && F'=\frac{1}{2}r^{D-2}p^2, \\
 && N'=0, \\ 
 && \varphi '=-Np,\\
 && (r^{D-2}p)'=0.
\end{eqnarray}
The solutions of the above equations are easily found 
\begin{eqnarray}
\label{ccsolution}
 && N=N_{\infty}, \\
 && p=\frac{Q}{r^{D-2}}, \\
\label{varphi}
 && \varphi=\frac{N_{\infty}Q}{(D-3)r^{D-3}} +\varphi _{\infty}, \\
\label{F}
 && F=-\frac{Q^2}{2(D-3)r^{D-3}} + \tilde{C}.
\end{eqnarray}
Here $\tilde {C}=M+\tilde{C}_0$, $\varphi_{\infty}$ is the value of $\varphi$ at
the infinity, which is conjugate to the  electric charge $Q$ of the solution, 
and the integration 
constant $N_{\infty}$ is the value of $N$ at the infinity which is conjugate to
the mass $M$. Therefore one can take $N_{\infty}=1$ by adjusting the time 
coordinate. In that case, we have solutions 
\begin{equation}
\label{csolution}
ds^2=-g^2(r)dt^2 +g^{-2}(r)dr^2 +r^2d\Sigma^2_{D-2},
\end{equation}
where
\begin{equation}
\label{cg^2}
g^2(r)=\left \{
\begin{array}{ll}
k+\left(\frac{r}{l}\right)^2 -\left[ \frac{2M+2\tilde{C}_0}{r} -
      \frac{Q^2}{(D-3)r^{D-2}}\right]^{\frac{1}{n-1}}      & \ \ {\rm for}\ \
            D=2n, \\
k +\left(\frac{r}{l}\right)^2 -\left[M+\tilde{C}_0-\frac{Q^2}{2(D-3)r^{D-3}}
       \right] ^{\frac{1}{n-1}}     &\ \ {\rm for}\ \ D=2n-1.
\end{array}\right.
\end{equation} 
Once again, the additive constant $\tilde{C}_0$ determines the ground states of
the solutions and can be fixed as in the previous section.
 When $D=3$, $\varphi$ and $F$ in (\ref{varphi}) and (\ref{F})
 should be replaced by 
\begin{eqnarray}
&& \varphi =-N_{\infty}Q\ln r +\varphi_0, \\
&& F=\frac{1}{2}Q^2\ln r +\tilde{C}
\end{eqnarray}
where $\varphi_o$ is an integration constant which is related to the choice 
 of zero electric potential. The metric function $g^2$ becomes
\begin{equation}
g^2(r)=-M +(r/l)^2-\frac{1}{2} Q^2 \ln r,
\end{equation}
which is just the charged BTZ black hole solution \cite{BTZ}.

We are not going to separately analyze here the solutions (\ref{cg^2}) for
 different curvature  $k$. Instead  we will give
 a unified description.  Just as the case of charged, spherically symmetric 
solutions \cite{BTZ2}, our solutions (\ref{cg^2}) may have two, one or no 
horizons depending on the relative value of the mass and
charge. Therefore the causal structure is similar to that of the
 Reissner-Nordstr\"om-anti-de Sitter black hole. But, we should notice 
that except for the singularity at $r=0$, the solutions (\ref{cg^2}) have 
another singularity at $r=r_c>0$ \cite{BTZ2} 
hidden by the black hole horizons $r_-$ and 
$r_+$: $0<r_c<r_-<r_+$, which can be see from the following curvature scalar
of the solutions
\begin{equation}
R=-(g^2)'' -\frac{2(D-2)}{r}(g^2)' +\frac{(D-2)(D-3)}{r^2}(k-g^2).
\end{equation}

In general, it is difficult to get an explicit expression of horizon of black 
hole solutions (\ref{cg^2}) in terms of the mass and charge. However, we 
can obtain an expression of mass in terms of the horizon
$r_+$ and charge. From (\ref{cg^2}), we have
\begin{equation}
\label{cmass}
M+\tilde{C}_0 =\left \{
\begin{array}{ll}
\frac{r_+}{2}\left [\left( k+\left(\frac{r_+}{l}\right)^2 \right)^{n-1}+\frac{Q^2}
     {(D-3)r^{D-2}_+}\right] & \ \ {\rm for}\ \ D=2n, \\
\left [k+\left(\frac{r_+}{l}\right)^2\right]^{n-1} +\frac{Q^2}{2(D-3)r^{D-3}_+}
             & \ \ {\rm for}\ \ D=2n-1.
\end{array} \right.
\end{equation}
According to the formula (\ref{tem}), the Hawking temperatures are 
\begin{equation}
\label{ctemp}
T=\left \{
\begin{array}{ll}
\frac{[k+(r_+/l)^2]^{2-n}}{2\pi (n-1)r_+}\left \{\frac{1}{2}\left [k+
    \left(\frac{r_+}{l}\right)^2\right]^{n-2} 
    \left [k+(2n-1)\left(\frac{r_+}{l}\right)^2 \right] -\frac{Q^2}{2r_+^{D-2}}
       \right \}
         & \ \ {\rm for}\ \ D=2n,\\
 \frac{[k+(r_+/l)^2]^{2-n}}{4\pi (n-1)}\left \{ \frac{2(n-1)r_+}{l^2} 
      \left [k+\left(\frac{r_+}{l}\right)^2\right]^{n-2} -\frac{Q^2}{2r_+^{D-2}}
        \right \}
        & \ \ {\rm for}\ \ D=2n-1.
\end{array} \right.
\end{equation}
Using (\ref{cmass}), one has
\begin{equation}
\label{cmassd}
\left(\frac{\partial M}{\partial r_+}\right)_{Q} =\left \{
\begin{array}{ll}
\frac{1}{2}\left [k+\left(\frac{r_+}{l}\right)^2\right]^{n-2}\left[k 
        +(2n-1)\left(\frac{r_+}{l}\right)^2\right]-\frac{Q^2}{2r_+^{D-2}}
         & \ \ {\rm for}\ \ D=2n, \\
\frac{2(n-1)r_+}{l^2}\left [k+\left(\frac{r_+}{l}\right)^2 \right]^{n-2}
     -\frac{Q^2}{2r_+^{D-2}} & \ \ {\rm for}\ \ D=2n-1.
\end{array} \right.
\end{equation}
Substituting (\ref{cmassd}) and (\ref{ctemp}) into (\ref{entropy}),
we find the entropy  
\begin{eqnarray}
\label{centropy1}
S&=&2\pi (n-1) \int^{r_+}_{0} r_+\left[k+\left (\frac{r_+}{l}\right)^2 \right 
     ]^{n-2}dr_+ \nonumber \\
 &=& \pi l^2 \left\{\left[ k+\left(\frac{r_+}{l}\right)^2\right]^{n-1}
       -k\right\},
\end{eqnarray}
for the even dimensional black holes, and 
\begin{eqnarray}
\label{centropy2}
S &=& 4\pi (n-1) \int^{r_+}_{0}\left [k+\left(\frac{r_+}{l}\right)^2\right]^{n-2}
       dr_+ \nonumber \\
&=& 4\pi (n-1) l\sum^{n-2}_{m=0}\left (
\begin{array}{c}
 n-2 \\ m
\end{array} \right) \frac{1}{2m+1}\left (\frac{r_+}{l}\right)^{2m+1} k^{n-2-m},
\end{eqnarray}
for the odd dimensional black holes. When $k=1$, the entropies (\ref{centropy1})
and (\ref{centropy2}) are totally the same as  those derived through the 
Hamiltonian analysis \cite{BTZ2}. The method used here seems to be  simpler.
Furthermore, it is found that the entropy formulas (\ref{centropy1}) and 
(\ref{centropy2}) are also exactly same as (\ref{entropytot1}) and 
(\ref{entropytot2}), which are derived as the charges are absent.  
 It verifies in some sense that the black 
hole entropy comes from a surface term of gravitational action at the 
horizon \cite{Cai4,Jaco,BTZ3}. That is,  the black hole entropy is  not
explicitly  related to the Lagrangian of matters.

Finally, we give the heat capacity of charged black holes
\begin{equation}
\label{cc1}
C_Q =\frac{2\pi (n-1)r_+^2 \left [ \triangle ^{n-2}\left[ (2n-1)
      \left(\frac{r_+}{l}\right)^2 +k\right] 
     -\frac{Q^2}{r_+^{D-2}}\right]}
     {\left [(2n-1)\left(\frac{r_+}{l}\right)^2-k  \right] 
      +\frac{Q^2 \triangle ^{1-n}}{r_+^{D-2}}
      \left[ (2n-1)k +(4n-5)\left(\frac{r_+}{l}\right)^2\right]}
\end{equation}
in the even dimensions;
\begin{equation}
\label{cc2}
C_Q=\frac{4\pi (n-1)r_+ \left[\frac{4}{l^2}\triangle ^{n-2} -
        \frac{Q^2}{(n-1)r_+^{D-1}} \right ]}
      {\frac{4}{l^2} +\frac{Q^2 \triangle ^{1-n}}{(n-1)r_+^{D-1}}
       \left[ (2n-3)k +(4n-7)\left(\frac{r_+}{l}\right)^2\right] }
\end{equation}
in the odd dimensions, where $\triangle =k +(r_+/l)^2 $. The behavior of heat
capacity is quite interesting.  Inspecting (\ref{cc2}), for the odd 
dimensional black holes, one can see that $C_Q$ is always positive and finite.
And $ C_Q=0$ when $4\triangle ^{n-2}/l^2 = Q^2/(n-1)r_+^{D-1}$, which 
corresponds to the extremal black holes, thereby the Hawking temperature 
(\ref{ctemp}) vanishes. For the even dimensional black holes, when $k=1$, 
the heat capacity has been analyzed in \cite{MP}. 
The heat capacity may be positive and negative, between them the heat capacity
has an infinite discontinuity. In the physical parameter regime, there are
three possibilities: the heat capacity has two, one and no infinite 
discontinuity(ies). When $k=0$ and $k=-1$, we find that the heat capacity 
(\ref{cc1}) is always positive and finite. When $T=0$, $C_Q=0$. This is the case
of extremal black holes.

\section{Conclusions }

In this work we have investigated the topological black holes in the 
dimensionally continued gravity which is a special class of the 
Lovelock gravity.   This is achieved by embedding the Lorentz group $SO(D-1,1)$ 
into the anti-de Sitter group $SO(D-1,2)$.
In this way the Lovelock gravity is divided into two branches depending on the 
spacetime dimension: even and odd dimensional cases.  The action is, in the odd 
dimensions, the Chern-Simons form for the anti-de Sitter group and, in the even 
dimensions, the Euler density constructed with the Lorentz part of the anti-de
Sitter curvature tensor. In the cases $D=3$ and $D=4$, the  two actions reduce
to the Einstein-Hilbert action with a negative cosmological constant
 in the general
relativity.  The Lovelock coefficients are
reduced to two parameters: cosmological constant and gravitational 
constant. The 
horizons of these topological black holes are $(D-2)$-dimensional 
hypersurfaces with
constant positive, zero or negative curvature $(D-2)k$. 
Therefore, the even horizon of 
black holes is no longer the $(D-2)$-dimensional
 sphere $S^{D-2}$. The horizons may
 also be toroidal or higher genus Riemann surfaces.

We have studied the three kinds 
of black holes and discussed their thermodynamic properties.  From the first 
law of thermodynamics of black holes, we have calculated these topological 
black hole entropy. It turns out that the entropy does not obey the usual 
area formula. When $k=1$, it reduces to that derived through a Hamiltonian 
analysis for the spherically symmetric black holes \cite{BTZ2}. 
Due to the different topological structures, these black holes 
manifest different thermodynamic behaviors. In the case $k=1$, the vacuum 
state is the anti-de Sitter
space in the even dimensions and in the odd dimensions, is zero mass black hole 
solution, as happens in the BTZ solution \cite{BTZ,Henx}, for which the horizon
and singularity coincide with each other at $r=0$. The heat capacity is alway 
positive for the odd dimensional black holes, but for the even dimensional black
holes, positive as $r_+>l/\sqrt{2n-1}$ and negative as $r_+<l/\sqrt{2n-1}$. That is,
it has a transition point at $r_+=l/\sqrt{2n-1}$, thereby the heat capacity suffers
from an infinite discontinuity.  In the case $k=0$, the vacuum state for both
solutions is the zero mass black hole as the zero mass BTZ solution. In this case,
the heat capacity is always positive for both solutions. In the case $k=-1$,
the vacuum state describes a black hole with horizon $r_+=l$ for both solutions.
This vacuum solution has some peculiar  properties. For the different dimensions,
the Hawking temperature is same as $T=1/2\pi l$, but the entropy has  different 
behaviors in the even dimensions and odd dimensions (\ref{zeroentropy}).
 This is 
because embedded in the different gravity, the same black hole solution
may have different entropy (Note that the vacuum solution in (\ref{In3}) when 
$k=-1$ and $M=Q=0$, has a entropy proportional to its area).
 In this case, the heat
capacity is also always positive for arbitrary dimensional black holes.
 In addition, the negative mass spectrum is allowed in the black hole solutions 
when $n=2\tilde {k}+1$, where $\tilde{k} \in {\cal Z}$. But, there exists a
 critical negative mass, beyond which the singularity at $r=0$ will be naked.
The critical mass has been found. Recently, the higher dimensional 
Chern-Simons supergarvity has been investigated \cite{TZ,Ban1}.
 It would be interesting
to study the supersymmetry of these topological black hole solutions found
in this paper and the constant curvature black holes \cite{Ban}

The charged topological black holes in this dimensionally continued 
gravity have been also considered.
The Hawking temperature, entropy and heat capacity have been calculated and 
analyzed. It has been found that for $k=0$ and $k=-1$ black hole solutions
(including those solutions (\ref{In3}) in the Einstein-Maxwell theory 
 see  the Appendix), the heat capacity is always positive, which means 
these black holes are more stable than the $k=1$ black holes. This work 
has extended the investigation on the static, spherically symmetric 
black holes in the dimensionally continued gravity \cite{BTZ2}.

\section*{Acknowledgments}

This work was supported by the Center for Theoretical Physics of Seoul National 
University.

\appendix

\section*{ Thermodynamics of higher dimensional topological black 
     holes in the Einstein-Maxwell theory}

In this appendix, we briefly discuss thermodynamics of the higher dimensional 
topological black holes (\ref{In3}) in the Einstein-Maxwell theory 
with a negative cosmological constant. For the discussion in the four 
dimensions see \cite{Brill}. For the solution (\ref{In3}), the vacuum
state is
\begin{equation}
\label{zeromass}
ds^2=-[k+(r/l)^2]dt^2 +[k+(r/l)^2]^{-1}dr^2 +r^2d\Sigma^2_{D-2},
\end{equation}
which is a $D$-dimensional  anti-de Sitter space with a $(D-2)$-dimensional 
hypersurface $\Sigma_{D-2}$ whose curvature is a constant $(D-2)k$. Therefore,
asymptotically,  the solution (\ref{In3}) is also locally isometric to 
the anti-de Sitter space. For the vacuum solution (\ref{zeromass}), 
as discussed in the text, the horizon is  absent as $k=1$, at $r_+=0$ 
coinciding with the singularity at $r=0$  when $k=0$, and is $r_+=l$ when
$k=-1$. Differing from the zero mass black hole (\ref{zero}) in the 
dimensionally continued gravity, the zero mass black hole (\ref{zeromass})
in the Einstein gravity obeys the area formula of entropy, which we will 
prove. 

 The solutions (\ref{In3}) may have two, one and no horizon(s). When 
the solutions describe black holes with non-degenerate horizon,
 using (\ref{tem}), we get the Hawking temperature in terms of the charge
and horizon radius $r_+$,
\begin{equation}
\label{atem}
T=\frac{1}{4\pi r_+}\left [(D-3)k +(D-1)\left(\frac{r_+}{l}\right)^2
  -\frac{16\pi^2 (D-3)Q^2}{\omega^2_{D-2} r^{2(D-3)}_+}\right].
\end{equation} 
 From the definition of horizon $g_{tt}(r_+)=0$, we have
\begin{equation}
\label{am}
\left (\frac{\partial M}{\partial r_+}\right)_Q =
 \frac{(D-2)\omega_{D-2}r_+^{D-4}}{16\pi} \left[ (D-3)k + 
     (D-1)\left(\frac{r_+}{l}\right)^2 -\frac{16\pi^2 (D-3)Q^2}
    {\omega^2_{D-2}r_+^{2(D-3)}}\right ].
\end{equation}                        
Substituting (\ref{atem}) and (\ref{am}) in to (\ref{entropy}), 
one has 
\begin{eqnarray}
S &=& \int ^{r_+}_0 T^{-1}\left(\frac{\partial M}{\partial r_+}\right)_Q dr_+ 
        \nonumber \\
  &=&\frac{(D-2)\omega _{D-2}}{4}\int ^{r_+}_0 r_+^{D-3}dr_+ 
      \nonumber \\
  &=&\frac{1}{4}\omega _{D-2}\ r_+^{D-2},
\end{eqnarray}
which is just one quarter of the horizon area. It also verifies that the
 black hole entropy in the Einstein gravity always satisfies the usual 
area formula, independent of  the topology of event horizon. One may wonder 
whether the entropy of $k=-1$ zero mass black holes (\ref{zero}) and
(\ref{zeromass}) obeys the entropy formula derived from (\ref{entropy}),
 because the formula (\ref{entropy}) seems non-applicable to these zero mass
black holes. Recall that the black hole entropy 
in fact comes from a surface term of the gravitational action under
 consideration, which is computed at the black hole horizon $r_+$ 
\cite{Cai4,Jaco,BTZ3}. Therefore, the black hole entropy is not related 
to whether the mass of black holes is zero or not. And hence the  result from
 (\ref{entropy}) is applicable  to the zero mass black holes.

The heat capacity of the black holes (\ref{In3}) is
\begin{equation}
\label{ac}
C_Q=\frac{(D-2)\omega_{D-2}r_+^{D-2}}{4}
     \frac{\left [ (D-3)k +(D-1)\left(\frac{r_+}{l}\right)^2 
     -\frac{16\pi^2(D-3)Q^2}{\omega^2_{D-2}r_+^{2(D-3)}}\right ] }
    {\left [-(D-3)k+ 3(D-1) \left(\frac{r_+}{l}\right)^2
      +\frac{16\pi^2 (D-3)(2D-5) Q^2}{\omega^2_{D-2}r_+^{2(D-3)}}\right]  }
\end{equation}
Inspecting the heat capacity, obviously, when $k=0$ 
and $k=-1$, one may find that 
it is always positive and finite. When $k=1$, it may be positive and negative,
between them an infinite discontinuity  occurs, which happens as the 
denominator of (\ref{ac}) vanishes. When the two horizons of black holes 
coincide with each other, we have $T=C_Q=0$, which corresponds to the 
extremal black holes. 

When $k=-1$, there is also the negative mass black holes. 
The critical mass, beyond which the singularity at $r=0$ will
be naked, is
\begin{equation}
M_{\rm c}=-\frac{(D-2)\omega_{D-2}r_c^{D-3}}{8\pi}
      \left[ 1-\frac{D-2}{D-3}\left( \frac{r_c}{l}\right)^2 \right ],
\end{equation}
where $r_c$ satisfies
\begin{equation}
r_c=l\sqrt{\frac{D-3}{D-1}}\left [1+\frac{16\pi ^2 Q^2}
        {\omega^2_{D-2}r_c^{2(D-3)}}\right]^{1/2}.
\end{equation}

\end{document}